\documentclass[conference]{IEEEtran}
\usepackage{cite}
\usepackage{amsmath,amssymb,amsfonts}
\usepackage{algorithmic}
\usepackage{graphicx}
\usepackage{textcomp}
\usepackage{xcolor}
\usepackage{booktabs}
\usepackage{threeparttable}
\usepackage{censor}

\def\BibTeX{{\rm B\kern-.05em{\sc i\kern-.025em b}\kern-.08em
    T\kern-.1667em\lower.7ex\hbox{E}\kern-.125emX}}
\begin{document}

\title{Reflective Homework as a Learning Tool: Evidence from Comparing Thirteen Years of Dual vs. Single Submission\\
\thanks{}
}

\author{
    \IEEEauthorblockN{Madhur Dixit, Kavya Lalbahadur Joshi, Kaveri Bhalchandra Konde, and Edward F. Gehringer}
    \IEEEauthorblockA{\textit{Department of Computer Science} \\\textit{North Carolina State University} \\
    Raleigh, USA \\
    \{mvdixit, kjoshi4, kkonde, efg\}@ncsu.edu}
}

\maketitle

\begin{abstract}
Dual-submission homework, where students submit work, receive feedback and then revise has gained attention as a way to foster reflection and discourage reliance on online answer repositories. This study analyzes 13 years of exam data from a computer architecture course to compare student performance under single versus dual-submission homework conditions. Using pooled $t$-tests on matched exam questions, we found that dual-submission significantly improved outcomes in a majority of cases. The results suggest that reflective resubmission can meaningfully enhance learning and may serve as a useful strategy in today’s AI-influenced academic environment. This full research paper also discusses pedagogical implications and study limitations.
\end{abstract}


\section{Introduction}
Dual-submission homework, where students submit assignments twice, first for an initial grade and feedback, then again after reflection, has been proposed as a way to foster metacognitive skills that support deeper learning. By encouraging students to actively review and correct their own work, this approach promotes greater engagement with course material and helps solidify conceptual understanding. It may also reduce opportunities for students to rely on external solution sources, such as online repositories, by requiring authentic revision. As these tools become more prevalent, dual submission introduces a structured pause for reflection that reinforces meaningful learning.

The present study statistically examines whether dual-submission homework has a measurable impact on student learning outcomes. Drawing on 13 years of data from a computer architecture course, we compare exam performance across semesters where students prepared using single-submission or dual-submission homework. Because many exam questions were reused across years, we are able to directly compare outcomes for students who had different homework formats but answered the same questions. The goal is to assess whether dual-submission preparation leads to any consistent patterns in student performance.

\subsection*{Research Questions}

This study is guided by the following research questions:

\begin{itemize}
    \item \textbf{RQ1:} Does dual-submission homework lead to improved student performance on course exams compared to single-submission homework?

    \item \textbf{RQ2:} Are performance improvements associated with dual-submission homework observable when controlling for question content and difficulty across multiple semesters?
\end{itemize}

\section{Theoretical Framework}
This study is grounded in two well-established learning theories: \textit{metacognition} and \textit{self-regulated learning} (SRL). Metacognition refers to the awareness and control of one's own thinking and learning processes. It includes activities such as evaluating one's understanding, identifying errors, and adapting learning strategies. Dual-submission homework directly engages these processes by requiring students to revisit their work, interpret instructor feedback, and revise their solutions.

Closely related is the theory of self-regulated learning, which emphasizes a learner’s ability to plan, monitor, and reflect on their own learning. According to SRL frameworks, students improve their performance when they receive feedback and are given opportunities to act on it. The second submission in dual-submission homework operationalizes this process, giving students a structured opportunity to self-correct and deepen their understanding.

By incorporating these principles, dual-submission homework serves not only as an assessment method but also as a reflective learning tool. This theoretical lens supports the hypothesis that students who engage in structured reflection will perform better on subsequent assessments, particularly those requiring conceptual reasoning.

\section{Related Work}
Over the past decade, the dual-submission homework methodology has emerged as a prominent strategy in engineering education.  Originally aiming to foster metacognitive skills and enhance student engagement, this method typically involves an initial submission of answers. After receiving the solutions from the instructor, students engage in a reflective or evaluative activity before submitting a revised version. The final grade often places substantial weight on this reflective submission.  Academic integrity has become an increasingly important motivation for the dual-submission approach, since a student who simply copies answers off the web will have difficulty explaining their reasoning or improving their answers.

The earliest paper of which we are aware is by Castellanos and Enszer from 2013 \cite{Castellanos2013}; the most recent are by Wood and Laughton \cite{Wood2023}, and Hume et al. \cite{Hume2023} from 2023.  The approaches share many features in common, though they differ in particulars.

\textit{Initial submission:}
Initial submissions universally involve students providing answers to homework problems. Certain approaches extend this phase by requiring additional elements, such as supporting work \cite{Breid2020}, methodology discussion \cite{DeGoede2020}, or reflective insights on confusion or learning challenges \cite{Goldberg2015, Castellanos2013}.  Wood et al.\@\cite{Wood2020} includes a cover sheet designed to prompt students' self-awareness early in the process, while Hume et al.\@\cite{Hume2023} emphasize self-revised problem sets, combining initial submissions with subsequent student-driven corrections.

\textit{First feedback:}
Feedback often includes providing solutions to the initial submissions \cite{Kearsley2016, Breid2019, Linford2020, Wood2020, DeGoede2020, Lund2020}, although grading approaches vary. Commonly, answers are assessed for completion \cite{Lura2015, Kearsley2016, Breid2019}, or effort \cite{Mota2019, Lund2020} rather than correctness. Breid et al.\@\cite{Breid2020} employ automated scoring, while DeGoede \cite{DeGoede2020} requires students to post explanations and interact with peers via a graded discussion board, followed by class discussions. Hume et al.\@\cite{Hume2023} propose a weighted grading model that prioritizes points for the initial submission while allocating a smaller percentage to revisions.

\textit{Second submission:}
The second submission phase generally focuses on metacognitive activities, such as corrections \cite{Goldberg2014, Kearsley2016, Chen2016, Mota2019, Breid2020, Wood2020} , self-grading \cite{Castellanos2013, Linford2020}, or reflections on the solution process \cite{Castellanos2013, Goldberg2014, Goldberg2015, Breid2019}. Homework wrappers, inspired by exam wrappers, serve as debriefing tools for students to assess their study strategies and effectiveness \cite{Chew2016, Chen2016, Lund2020}. Notably, Chew et al.\cite{Chew2016} offer empirical evidence on the use of homework and exam wrappers in engineering contexts, showing improvements in students’ self-regulated learning and performance in statics problem-solving. \cite{Lund2020} However, student compliance can vary, as noted by Goldberg \cite{Goldberg2014}, where substantial non-completion of reflection assignments prompted in-class revisions in her follow-up paper \cite{Goldberg2015}, achieving higher completion rates.

Variations include collaborative approaches by DeGoede \cite{DeGoede2020}, where students evaluate peers’ problem-solving methods on discussion boards, and Mota \cite{Mota2019}, which incorporates team discussions before individual corrections. Lura et al.\@\cite{Lura2015} integrate quizzes with altered homework problems in the second round, while Chen \cite{Chen2016} offers a multi-round system including quiz corrections. Wood and Laughton \cite{Wood2023} synthesize best practices across implementations, proposing graduated responsibility for self-directed learning. For sophomore-level courses, dual submissions include both corrections and reflections for completion credit, while junior- and senior-level courses scaffold increasing responsibility for self-assessment and problem-solving accuracy.

\textit{Final grade:}
Grading schemes among dual-submission methodologies are fairly diverse. Metacognitive-focused approaches often include correctness-based grading components \cite{Castellanos2013, Goldberg2014, Chew2016}. Completion grading, emphasizing effort or thoroughness, appears in a few strategies \cite{Breid2020, Wood2020}. Linford \cite{Linford2020} at USMA combines student self-assessments with instructor evaluations to determine a final grade. DeGoede \cite{DeGoede2020} grades engagement and skill development via discussion posts. Strategies involving quizzes \cite{Lura2015, Chen2016} base final grades on quiz scores, with Chen \cite{Chen2016} noting limited success with quiz corrections.

Hume et al.\@\cite{Hume2023} highlight the neutral-to-positive outcomes of self-revised problem sets, with increased engagement by struggling students and favorable instructor perceptions regarding grading ease. Wood and Laughton \cite{Wood2023} emphasize the importance of aligning dual-submission methods with scaffolded learning goals at various academic levels, ensuring students experience authentic, reflective problem-solving akin to professional engineering practice.

In summary, dual-submission homework methodologies exhibit a multifaceted framework for integrating metacognitive practices into engineering education. By blending individual reflection, collaborative elements, and innovative feedback mechanisms, these approaches provide scalable solutions for fostering deeper learning and adapting to diverse institutional needs. However, while prior studies provide valuable insights into the design and implementation of dual‑submission systems, most focus on student perceptions, small-scale case studies, or short-term interventions. Few offer long-term, data-driven evidence of exam performance outcomes. One notable exception is the study by Jay and Dodd \cite{jay2022efficacy}, which provides quantitative exam-based evidence, though limited in scope. Building on this, this paper analyzes 13 years of matched exam question data, offering a large-scale, longitudinal evaluation of dual‑submission effectiveness. This study builds on previous work by offering a longitudinal, question-level comparison of exam performance across dual- and single-submission cohorts in an engineering course.

\section{Methodology}
The data for this study were obtained from CSC/ECE 506, Architecture of Parallel Computers, which has been taught once or twice a year by the last author over the period from 2012--2025, typically face-to-face in the spring, with an online section in the summer. Over this period, the textbook has remained constant (except for a new edition in 2016), as has most of the material covered in class.  The course has two midterm exams and one final exam.  Each exam consists of six questions worth 20 points each, with the lowest of the six question scores being dropped.  Questions from one semester are often reused in later years.  Often this is done by changing the numbers or the reference trace (the input string).  The exact question may also be reused, but not within four years of the previous usage.  This practice was informed by a previous FIE paper \cite{Gehringer2004}, which did not find any academic-integrity compromise from reusing test questions more than two years apart.

In recent years, three problem sets have been assigned during the course, one before each of the three exams. The material covered on each problem set is similar to the corresponding exam material, although it is rare for exactly the same type of problem to be found on both a problem set and an exam.  In the ensuing discussion, it will be said that students had ``prepared for'' a particular exam by working a particular problem set.

The author became aware of the dual-submission homework approach at an ASEE regional conference, and saw further presentations on the topic at ASEE annual conferences.  He decided to team up with the authors of one of those papers \cite{jay2022efficacy} in an effort to show that dual submission held a benefit for students.  Beginning in 2023, he began using the dual-submission approach.  It was used
\begin{itemize}
    \item for one problem set in Spring 2023,
    \item for two problem sets in Spring 2024,
    \item for three problem sets in Summer 2024 and Spring 2025.
\end{itemize}
Consequently, it can be said that students prepared for the corresponding number of exams using dual-submission homework.

\subsection{Data Collection}
Over this 13-year period, a total of 318 questions were asked on exams.  Questions were categorized based on whether students prepared using single-submission or dual-submission homework. As it turns out, 33 distinct questions were used both on at least one exam where students had done single-submission homework and also on at least one exam where they prepared with dual-submission homework. This allows us to compare the performance by students who prepared for that question by single- and dual-submission homework.

The author attempts to track scores for each individual question on each test.  But sometimes the data was not collected, or was lost.  If we did not have data for at least one single-submission usage of the question and one dual-submission usage, the question was removed from the dataset.  This resulted in the removal of Questions 3, 4, 9, and 20.  The results are based on the remaining 29 questions.

\subsection{Statistical Analysis}
For each of the 29 questions, the number of students responding ($N$) and the mean score they received ($\mu$) on that question. The $N$ values were used as weights when calculating aggregated means and variances. Welch's pooled $t$-tests were conducted using a custom Python script (\texttt{code.py})~\cite{analysisScript} to compare the scores of students who prepared using single- versus dual-submission homework. This test was chosen because it does not assume equal variances between groups and is appropriate when comparing means across samples with unequal sizes and heteroscedasticity—conditions present across many of the matched question groups in the dataset. The resulting $p$-values from these comparisons are summarized in Table~II.
.

\section{Results}
\subsection{Summary of Key Statistical Results}
Table I lists a subset of the data used in our analysis, showing five selected questions along with the semesters in which they appeared, the mode of homework preparation, and the corresponding student performance. While these questions were not chosen to be representative in a statistical sense, they serve to illustrate the structure and nature of the full dataset. Specifically, they provide the reader with a concrete view of how question-level data was organized and recorded. The complete single- versus dual-submission samples used for the $t$-tests were constructed from the full dataset, which includes all 29 questions meeting the inclusion criteria.

\begin{table}[htbp]
\vspace{12pt}
\caption{\textsc{\fontsize{8}{10}\selectfont Scores for Sample Questions, by Semester \& Preparation Type}}
\vspace{6pt}
\begin{center}
\begin{threeparttable}
\begin{tabular}{|c|p{2.3cm}|c|c|c|c|}
\toprule
\textbf{Ques.\#} & \textbf{Topic} & \textbf{Semester} & \textbf{Grp.} & \textbf{N} & $\mu$ \\
\hline
1 & Amdahl's Law & Sum 22 & Single & 6  & 17.83 \\
1 & Amdahl's Law & Sum 23 & Single & 10 & 19.10 \\
1 & Amdahl's Law & Sum 24 & Dual & 12 & 18.83 \\
1 & Amdahl's Law & Spring 21    & Single & 37 & 18.46 \\
1 & Amdahl's Law & Spring 25    & Dual & 41 & 17.51 \\
\hline
10 & Dragon protocol & Spring 14    & Single & 33 & 14.64 \\
10 & Dragon protocol & Sum 22  & Single & 6  & 19.66 \\
10 & Dragon protocol & Spring 23    & Dual & 32 & 19.00 \\
10 & Dragon protocol & Spring 24    & Dual & 45 & 15.76 \\
\hline
11 & DSM coherence & Sum 23 & Single & 9  & 18.89 \\
11 & DSM coherence & Spring 24   & Dual & 43 & 17.05 \\
11 & DSM coherence & Spring 22   & Single & 52 & 18.08 \\
\hline
15 & Linked data structures & Fall 16   & Single & 110 & 17.04 \\
15 & Linked data structures & Spring 25   & Dual & 40  & 15.58 \\
15 & Linked data structures & Spring 20   & Single & 36  & 16.52 \\
15 & Linked data structures & Sum 24 & Dual & 12  & 13.67 \\
\hline
31 & Update vs. invalidation protocols  & Spring 12   & Single & 38 & 13.94 \\
31 & Update vs. invalidation protocols  & Spring 20   & Single & 32 & 15.62 \\
31 & Update vs. invalidation protocols  & Spring 25   & Dual & 42 & 15.14 \\
31 & Update vs. invalidation protocols & Sum 24 & Dual & 11 & 14.82 \\
\bottomrule
\end{tabular}
\end{threeparttable}
\end{center}
\end{table}

Table II summarizes the pooled $t$-test results:
\cite{resultsData}

\begin{table}[htbp]
\vspace{12pt} 
\caption{\textsc{\fontsize{8}{10}\selectfont Summary Statistics from Pooled $t$-Tests}.}
\vspace{6pt} 
\begin{center}
\begin{threeparttable}
\begin{tabular}{|cccccc|}
\toprule
\textbf{Ques.\#} & \textbf{$N_\mathrm{single}$} & \textbf{$N_\mathrm{dual}$} & \textbf{$\mu_\mathrm{single}$} & \textbf{$\mu_\mathrm{dual}$} & \textbf{$p$-value} \\
\hline
1 & 53 & 53 & 18.51 & 17.81 & 0.364 \\
2 & 165 & 11 & 14.56 & \textcolor{green!90!black}{18.45} & $2.7 \times 10^{-6}$ \\
5 & 166 & 45 & 14.08 & 15.22 & 0.0943 \\
\hline
6 & 5 & 32 & 15.80 & 12.66 & 0.055 \\
7 & 343 & 39 & \textcolor{green!90!black}{11.89} & 9.34 & 0.013 \\
8 & 37 & 41 & 9.62 & 9.85 & 0.796 \\
\hline
10 & 39 & 77 & 15.41 & \textcolor{green!90!black}{17.11} & 0.019 \\
11 & 61 & 43 & \textcolor{green!90!black}{18.20} & 17.05 & 0.030 \\
12 & 10 & 43 & 10.30 & \textcolor{green!90!black}{13.74} & 0.032 \\
\hline
13 & 136 & 11 & 15.53 & 14.82 & 0.709 \\
14 & 51 & 54 & 15.51 & \textcolor{green!90!black}{17.26} & 0.002 \\
15 & 146 & 52 & \textcolor{green!90!black}{16.91} & 15.14 & 0.001 \\
\hline
16 & 33 & 11 & 19.33 & 18.73 & 0.426 \\
17 & 73 & 10 & 12.00 & 10.60 & 0.553 \\
18 & 9 & 45 & 17.22 & 16.18 & 0.170 \\
\hline
19 & 173 & 13 & 15.79 & 15.73 & 0.952 \\
21 & 6 & 38 & 15.33 & 13.06 & 0.174 \\
22 & 9 & 45 & 14.00 & 14.64 & 0.736 \\
\hline
23 & 33 & 13 & 8.60 & \textcolor{green!90!black}{18.60} & 0.000 \\
24 & 69 & 43 & 15.97 & 15.68 & 0.660 \\
25 & 33 & 11 & \textcolor{green!90!black}{18.06} & 14.91 & 0.034 \\
\hline
26 & 10 & 52 & 15.50 & 16.38 & 0.572 \\
27 & 38 & 45 & 13.45 & \textcolor{green!90!black}{15.98} & 0.010 \\ 
28 & 160 & 12 & \textcolor{green!90!black}{18.34} & 13.00 & 0.038 \\
\hline
29 & 33 & 11 & 14.60 & 16.36 & 0.089 \\
30 & 51 & 46 & 16.18 & 16.78 & 0.565 \\
31 & 108 & 53 & 15.49 & 15.07 & 0.590 \\
\hline
32 & 340 & 12 & 11.43 & \textcolor{green!90!black}{14.83} & 0.010 \\
33 & 44 & 44 & 14.36 & \textcolor{green!90!black}{15.36} & 0.028 \\
\bottomrule
\end{tabular}
\begin{tablenotes}
\footnotesize
\item[] Green-highlighted values indicate significant difference ($p < 0.05$), favoring the respective group.
\end{tablenotes}
\end{threeparttable}
\end{center}
\end{table}

\begin{figure*}[htbp]
\centering
\includegraphics[width=\textwidth]{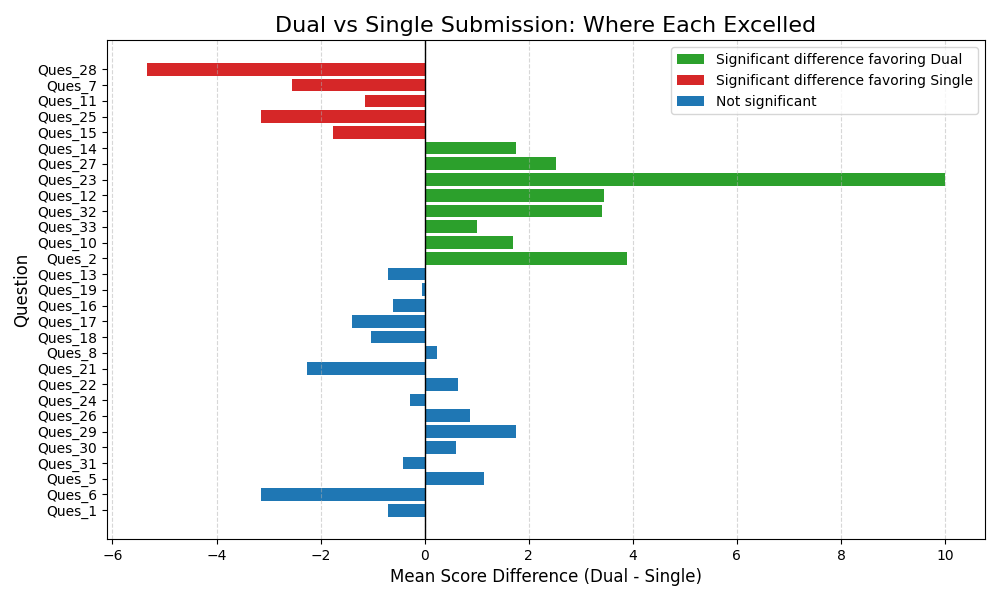}
\caption{Mean Score Comparison for Each Question}
\label{fig:mean_scores}
\end{figure*}

\subsection{Expanded Analysis of Results}
Out of 33 question groups analyzed, 13 displayed statistically significant differences ($p < 0.05$). Among these, 8 favored dual-submission preparation, 5 favored single-submission. Notable examples:
\begin{itemize}
    \item \textbf{Question 10}: Dual submission group ($N = 77$) had a mean score of 17.11 compared to 15.41 for single ($N = 39$), with $p = 0.019$, supporting dual preparation.
    \item \textbf{Question  12}: A large difference was observed where the dual group ($N = 43$) scored 13.74 while the single group ($N = 10$) scored 10.30, $p = 0.032$.
    \item \textbf{Question 11}: Interestingly, this question favored single submission with a mean of 18.20 ($N = 61$) compared to 17.05 ($N = 43$), $p = 0.031$.
\end{itemize}

The average pooled variance among dual-submission cases was higher (mean: 20.01) than single-submission cases (mean: 16.55), suggesting a wider distribution of performance—potentially due to the reflective learning curve introduced by dual submission. The average degrees of freedom across all Welch $t$-tests was approximately 44.8, indicating a reasonable approximation in pooled tests.

To better visualize the findings, Figure 1 plots mean scores of all 29 usable questions, showing clusters where dual or single preparation excelled.


Overall, while not all results were significant, the predominance of dual-submission favorability suggests its effectiveness, particularly when the questions require deeper conceptual application.

To get an idea of whether dual submission was better overall than single submission, we took the weighted harmonic mean of the $p$-values from questions favoring single submission, and compared it with its counterpart for the questions favoring dual submission, as elaborated below. 

\subsection{Aggregate Considerations}
The question-level data suggests a broader trend toward dual-submission efficacy. An aggregated statistical approach could further quantify this cumulative impact. Additionally, post-ChatGPT introduction data raises important considerations for the effectiveness of dual-submission methods compared to traditional single submissions.

\subsection{Weighted Harmonic Mean of Significant $p$-Values}

To better understand the aggregate strength of the significant findings, we computed the weighted harmonic mean of all $p$-values that were statistically significant ($p < 0.05$). This measure helps summarize the collective evidence favoring dual-submission over single-submission preparation, while appropriately giving more weight to comparisons based on larger samples (as reflected in their degrees of freedom).

Since one $p$-value was reported as zero, we substituted  a near-zero value ($1 \times 10^{-10}$) for it to avoid division errors. The harmonic-mean formula was:
\[
H = \frac{\sum w_i}{\sum \left( \frac{w_i}{p_i} \right)}
\]
where $w_i$ represents the degrees of freedom for test $i$, and $p_i$ is the corresponding $p$-value.

The harmonic mean of the eight questions for which dual submission outperformed single submission is $1.09 \times 10^{-9}$ (the harmonic mean of the seven questions with a non-zero $p$-value is $3.64 \times 10^{-6}$).  Contrast this with the harmonic mean of the five questions where single submission outperformed, which is $2.49 \times 10^{-3}$. This illustrates that there is stronger evidence supporting the outperformance of dual-submission preparation. 

Table II provides the data used for these computations.

\section{Discussion}

The weight of the evidence strongly supports the effectiveness of dual-submission homework as a learning strategy. Among the 13 exam questions with significant $p$-values, 8 favored dual submission, and the harmonic mean of the $p$-values was substantially lower for those items. This indicates that not only is performance more frequently better with dual submission, but also that the performance gap is more pronounced in its favor. Taken together, these results suggest that students in this cohort achieved greater learning through the dual-submission approach.

This improvement can be explained by established cognitive principles. Submitting work twice improves learning because when students resubmit, they're essentially practicing retrieval practice (recalling information) and error correction, both of which significantly strengthen memory and understanding.

The act of a second submission prompts students to actively reflect on feedback and fix their mistakes. This not only reinforces concepts but also encourages self-regulated learning, where students take more ownership of their educational journey. This approach also reduces cognitive load and performance pressure, helping students focus on mastering the material rather than just getting the right answer the first time.

These findings are consistent with prior educational research, which has shown that structured revision cycles, active correction, and feedback loops support deeper learning. In contrast to peer-review or single-pass methods, dual submissions offer instructor-guided feedback and a second chance to apply it, leading to better outcomes on objective assessments.

Many papers on dual submission tout its benefits: improved metacognition, academic integrity, time efficiency for instructors, and reduced student stress. But this is only the second paper (after Jay and Dodd \cite{jay2022efficacy}) to demonstrate a measurable learning advantage. This is especially notable because the dual-submission semesters in this study coincide with the emergence of tools like ChatGPT. Despite the ease of generating initial answers using AI, the second reflective submission appears to act as a pedagogical safeguard requiring meaningful re-engagement with the material and ensuring authentic learning.\footnote{For a discussion of the impact of generative AI on dual-submission homework, see the discussion by Gehringer et al. \cite{Gehringer2025}.}

From a pedagogical standpoint, dual submission can serve as a scalable, formative strategy. Instructors might consider extending it to labs, quizzes, or programming assignments any context where feedback and iteration can improve conceptual understanding and reduce grading-related anxiety.

\section{Threats to Validity}
Several limitations must be acknowledged. Data were sourced from a single course, which limits generalizability. Although the instructor was the same, the material was the same, and the textbook was the same (except for edition), changes in student demographics and program selectivity over 13 years may have influenced the outcomes. Additionally, ``preparation" encompasses a broad range of activities, and variations in grading rigor among different graders could introduce inconsistencies.

\section{Conclusion and Future Work}
This study hints at significant educational benefits from dual-submission homework through robust statistical analyses of exams over a 13-year period. Dual-submission assignments appear to encourage deeper student reflection, improve learning outcomes, and resist negative influences of online solution repositories. Future research should extend these findings to other courses and further investigate the impacts of emerging technologies such as artificial intelligence on homework strategies.

\bibliographystyle{IEEEtran}
\bibliography{dual}


\vspace{12pt}

\end{document}